\documentclass[prb,showpacs]{revtex4}
\usepackage[dvips]{graphicx}
\usepackage{amsmath,amssymb}
\usepackage{float}
\usepackage{bm}
\usepackage{subfigure}

{%
   \begin{list}{$\bullet$\ \ }
   {%
      \setlength{\itemindent}{0pt}
      \setlength{\leftmargin}{0.5zw}
      \setlength{\rightmargin}{0zw}
      \setlength{\labelsep}{0.5zw}
      \setlength{\labelwidth}{0zw}
      \setlength{\itemsep}{0em}
      \setlength{\parsep}{0em}
      \setlength{\listparindent}{0zw}
   }
}{%
   \end{list}%
}

\newcommand{\ignore}[1]{}
\newcommand{\beq}{\begin{equation}}
\newcommand{\eeq}{\end{equation}}

\begin{document}

\title
{Numerical study on ESR by making use of Wiener-Khinchin relation in time domain}

\author{Hiroki Ikeuchi$^{1,2}$}  
\email[Corresponding author. Email address: ]{ikeuchi@spin.phys.s.u-tokyo.ac.jp} 
\author{Sylvain Bertaina$^{3}$}  
\author{Seiji Miyashita$^{1,2}$}

\affiliation{$^{1}
${\it Department of Physics, Graduate School of Science,} The University of Tokyo, 7-3-1, Bunkyo-Ku, Tokyo, Japan \\
$^{2}${\it CREST, JST, 4-1-8 Honcho Kawaguchi, Saitama, 332-0012, Japan}\\
$^{3}${\it Aix-Marseille Universit\'e, CNRS, IM2NP UMR7334, F-13397 Marseille Cedex 20, France}
}

\date{\today}

\begin{abstract}  
To evaluate ESR spectrum at finite temperatures for specified spatial configurations of spins is very important issue to study quantum spin systems. Although a direct numerical estimation of the Kubo formula provides exact data, the application is limited to small size of the system because of the restriction of the computer capacity. The method of the Fourier transform of the autocorrelation function improved the restriction. As an extension of the method, 
we propose a new method for numerical calculation of the ESR spectrum from the time evolution of the magnetization by making use of the Wiener-Khinchin theorem.
\end{abstract}


\maketitle


\section{Introduction}

Quantum spin systems have attracted interests in the decades, because they exhibit various nontrivial characteristics. In particular, the strong quantum fluctuation and/or competition among the interaction (frustration) cause the singlet pair (the so-called valence bond) to behave as a unit of degree of freedom, and various novel concepts, e.g., the valence bond solid, resonating valence bond, and also magnon BEC, etc. have been developed. ESR is one of the major tools to study these properties. 
In particular, spatial arrangements of magnetic ions play important role for the property, and we need to treat the model microscopically.

Another important topic related to the finite size effect is the effect of nonmagnetic defects in quantum spin chains.  In spin $S=1/2$ antiferromagnetic Heisenberg chain, the quantum fluctuation prevents the system from being ordered even at $T=0$K. A nonmagnetic defect breaks the translational symmetry and polarizes the surrounding spins\cite{R1,R2}.  
Such a system is described by an open spin chain and some effects on susceptibility have been study on the Pd doped chain Sr$_2$CuO$_3$\cite{R3,R4}. Due to large anisotropy no ESR signal of such defects have been report. Only magnetic resonance of intrinsic defects in spin-Peierls CuGeO$_3$\cite{R5} is reported since the signal from the chain drops for $T<T_{\rm sp}$. 
Recently it has been reported in organic spin chain Fabre salt, the ESR signal of an intrinsic defect in Heisenberg antiferromagnetic chain\cite{R6}. Moreover, they observed the coherence signal of the correlated defect which could have an important impact in the domain of quantum information\cite{R7,R8,R9,R10}. 
The knowledge of the ESR of intrinsic defect in spin chain is an important problem and is limited by the number of spins of the open chain one can calculate.   

To study these aspects of quantum spin systems, the direct estimation of the ESR spectrum for given Hamiltonian has been investigated. The most simple way is to calculate the
Kubo formula directly by making use of the eigenvalues and eigenvectors obtained by diagonalization of the Hamiltonian\cite{miyashita,cepas}.  
But, the application of this method is limited to small systems for which we can obtain all the eigenvalues and their eigenvectors. For example, for the spin system of $N$ spins of $S=1/2$, we need the memory of $2^{2N}$. By making use of the symmetries of the system we may reduce the dimensions of the block diagonalized Hamiltonian, but still the size is limited in $N<20$.

 To relieve this restriction, a method to obtain the autocorrelation function by making use of time evolution of state has been introduced.\cite{iitaka,machida} 
The spectrum is obtained by Fourier transform of the autocorrelation function.
 In this method, 
The time evolution of autocorrelation function is obtained by the application of the 
time-evolution operator $e^{-i{\cal H}t}$ by making use of the Chebyshev iteration formula. Here we need the memory only of the order $2^N$, and we could study double size of the case of the diagonalization method. In principle, we need to take average over the thermal distribution of the initial state, but thanks to the idea of the thermal state\cite{iitaka,machida,hams,shimizu} we do not need to take the complete average. 
By this method, 
the thermal property of the ESR spectrum for the single molecular magnet V$_{15}$ which consists of 15 $S=1/2$ spins has been studied.\cite{machida}
In this method, the time evolution of the autocorrelation is obtained in a finite time domain $0<t<T$, and thus spectrum is suffered from the so-called Gibbs oscillation, and the prescription of gaussian masking is necessary to smear out the oscillation.

In the present paper, we propose an alternate method to obtain ESR spectrum from a time evolution of a magnetization by making use of
the Wiener-Khinchin theorem, which relates the spectrum density of magnetization dynamics and the Fourier transform of the autocorrelation function which gives the ESR spectrum.
In quantum system, the definition of the dynamics of magnetization is tricky and we give a quantum version of Wiener-Khinchin relation, i.e, an explicit relation between 
Fourier transform of the autocorrelation function and the spectrum density in the quantum case.  By making use of the relation we can obtain the ESR spectrum at finite temperatures. In this method the effect of the Gibbs oscillation is significantly
reduced.

The outline of this paper is as follows.
The methods previously used are explained in Sec~\ref{sec_method}. 
In Sec.~\ref{sec_WK}, we explained the new method motivated by the Wiener-Khinchin theory.
In Sec.~\ref{sec_summary}, we give the summary of the paper and discussion on related problems.
In Appendix A, we give description of a spin system for which we demonstrate the methods, and
in Appendix B, we explain the Gibbs oscillation.

\section{Previous Methods}\label{sec_method}

\subsection{Kubo formula and ESR spectrum} 
The ESR spectrum is obtained by the Kubo formula\cite{kubo_tomita,kubo}. 
The imaginary part of the dynamical susceptibility $\chi''(\omega)$ 
is given by
\beq
	\chi''(\omega)=
	\frac{1}{2}(1-\mathrm{e}^{-\beta\omega})\int_{-\infty}^{\infty}\langle M^{x}(0)M^{x}(t) \rangle_{\mathrm{eq}}\mathrm{e}^{-\mathrm{i}\omega t}\mathrm{d}t,
\label{eq:chi}
\eeq	
and The ESR absorption spectrum is given by
\beq
	I^{x}(\omega)=\frac{\omega\lambda_{0}^{2}}{2}\chi''(\omega).
\eeq
Here, we adopt usual notations:
\begin{eqnarray}
      	M^{x}(t)=\mathrm{e}^{\mathrm{i}\mathcal{H}t}M^{x}\mathrm{e}^{-\mathrm{i}\mathcal{H}t},\quad M^{x}=\sum_{i=1}^{N}S_{i}^{x},\\
	\langle \cdot \rangle_{\mathrm{eq}}=\mathrm{Tr}[\hspace{1mm}\cdot\hspace{1mm}\mathrm{e}^{-\beta\mathcal{H}}]/\mathrm{Tr}[\mathrm{e}^{-\beta\mathcal{H}}].
\end{eqnarray}

\subsection{Numerical methods} 
For numerical analysis of the ESR absorption spectrum, several methods have been developed. 
There are essentially two types of methods:
 (1) Exact diagonalization method\cite{miyashita,cepas}  and (2) Time-evolution of the autocorrelation function method\cite{iitaka}.
 
\subsubsection{Exact diagonalization method}

For direct estimation of the Kubo formula, we may explicitly evaluate the formula by making use of the set of the eigenvalues and the eigenvectors  $\{{{E_{n},|n\rangle}\}_{n=1}^{D}}$  of  the hamiltonian $\mathcal{H}$ where $D$ is the dimension of Hilbert space of ${\cal H}$:
\beq
\mathcal{H}|n\rangle=E_{n}|n\rangle
\eeq
obtained by a numerical diagonalization.
The autocorrelation function $\langle M^{x}(0)M^{x}(t)\rangle_{\mathrm{eq}}$ is expressed as
\begin{eqnarray}
	 \langle M^{x}(0)M^{x}(t)\rangle_{\mathrm{eq}}&=&\sum_{n}\langle n|M^{x}\mathrm{e}^{\mathrm{i}\mathcal{H}t}M^{x}\mathrm{e}^{-\mathrm{i}\mathcal{H}t-\beta\mathcal{H}}|n\rangle/Z\\
	&=&\sum_{n}\sum_{m,m'}\langle n|M^{x}|m\rangle\langle m|\mathrm{e}^{\mathrm{i}E_{m'}t}|m'\rangle\langle m'|M^{x}|n\rangle
\mathrm{e}^{-\mathrm{i}E_{n}t-\beta E_{n}}/Z\\
	&=&\sum_{m,n}|\langle m|M^{x}|n\rangle|^{2}\mathrm{e}^{\mathrm{i}(E_{m}-E_{n})t-\beta E_{n}}/Z,
\end{eqnarray}
where $Z$ is the partition function 
\beq
Z=\sum_{n}\mathrm{e}^{-\beta E_{n}}.
\eeq
The Fourier transform of this reads
\begin{eqnarray}
	\int_{-\infty}^{\infty}\langle M^{x}(0)M^{x}(t)\rangle_{\mathrm{eq}}\mathrm{e}^{-\mathrm{i}\omega t}\mathrm{d}t
	&=&\sum_{m,n}|\langle m|M^{x}|n\rangle|^{2}\mathrm{e}^{-\beta E_{n}}\int_{-\infty}^{\infty}\mathrm{e}^{-\mathrm{i}\left(\omega-(E_{m}-E_{n})\right)t}\mathrm{d}t/Z\\
	&=&\sum_{m,n}|\langle m|M^{x}|n\rangle|^{2}\mathrm{e}^{-\beta E_{n}}2\pi\delta\left(\omega-(E_{m}-E_{n})\right)/Z.
\end{eqnarray}
This yields the imaginary part of the dynamical susceptibility as
\begin{eqnarray}
	\chi''(\omega)&=&\frac{1-\mathrm{e}^{-\beta\omega}}{2}\sum_{m,n}|\langle m|M^{x}|n\rangle|^{2}\mathrm{e}^{-\beta E_{n}}2\pi\delta\left(\omega-(E_{m}-E_{n})\right)/Z\\
	&\equiv&\sum_{m,n}D_{m,n}\delta\left(\omega-\omega_{m,n}\right),
	\label{chidelta}
\end{eqnarray}
where
\begin{eqnarray}
	D_{m,n}\equiv\pi(\mathrm{e}^{-\beta E_{n}}-\mathrm{e}^{-\beta E_{m}})|\langle m|M^{x}|n\rangle|^{2}/Z,\quad \omega_{m,n}\equiv E_{m}-E_{n}.
\end{eqnarray}
We may treat only the range of $\omega_{m,n}>0$ since we are considering the absorption peak, not the emission.  Note that $\chi''(\omega)>0$ for $\omega>0$.

In this method, the spectrum is given by an ensemble of delta functions. Thus to draw the spectrum, 
we convert them in a continuous form. For example we may use bins in the $\omega$ axis, or
we replace the delta function by a gauss distribution with some variance which 
represents a finite resolution.

\begin{figure}[H]
	\begin{center}
	\includegraphics[width=100mm]{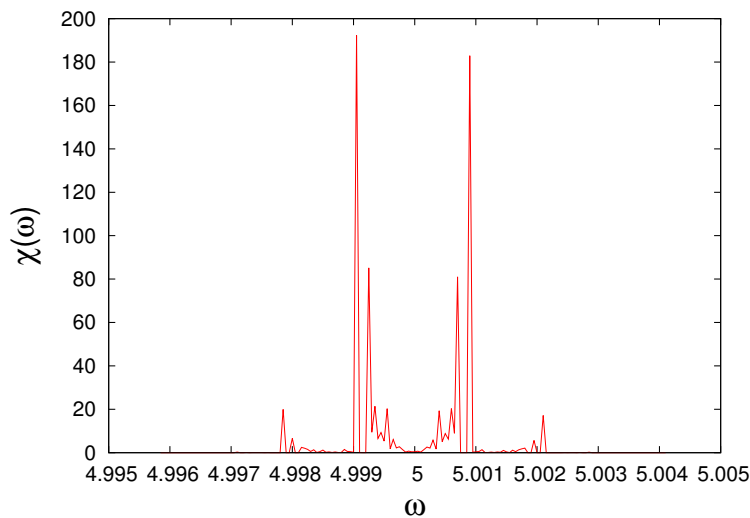}
	\end{center}
	\vspace{10mm}
	\caption{The spectrum of an antiferromagnetic Heisenberg chain obtained by exact diagonalization method: $N=14$, $T=100\mathrm{K}$, $H=5\mathrm{K}$, and $\Delta\omega=0.00005$.}
	\label{ED}
\end{figure}

 This method is exact, but we need to obtain all the eigenvalues and their eigenstates. Therefore, we need to store the matrix of the size $D$, which is $2^N$ for systems of $N$ spins with $S=1/2$. 
This requires $D^2$ in the memory. 
If the system has symmetry, we may reduce the size. For example if the system conserves $M^{z}$, then $D$ is reduced to $_NC_{N/2+M^z}$. Moreover for ESR only the uniform mode is relevant and thus only the fully symmetrized states are necessary, which also reduces $D$. 
However, the limitation of the memory prevents us from calculation more than $N=20$ for systems of $S=1/2$.
In Fig.~\ref{ED}, an example of spectrum obtained by exact diagonalization method for an antiferromagnetic Heisenberg chain with $N=14$ with the static field $H=5\mathrm{K}$ at the temperature $T=100\mathrm{K}$. The notation of the model is explained in Appendix A.
The data is given as ensemble of the delta peaks (Eq.(\ref{chidelta})), we made a histogram with small bins to show the spectrum with a mesh of the frequency $\Delta\omega=0.00005$.

\subsubsection{Time-evolution of the autocorrelation function method}
\label{sec:TDACFN}

As we find in the Kubo formula(Eq.(\ref{eq:chi})), the spectrum is given by the autocorrelation function.
Thus, if we have the autocorrelation function, the spectrum is given by its Fourier transform. 

First, we derive the explicit expression to calculate in the time-domain method. The integral is divided to two parts.
 \begin{eqnarray}
	\int_{-\infty}^{\infty}\langle M^{x}M^{x}(t)\rangle_{\mathrm{eq}}\mathrm{e}^{-\mathrm{i}\omega t}\mathrm{d}t
&=&\int_{0}^{\infty}\langle M^{x}M^{x}(t)\rangle_{\mathrm{eq}}\mathrm{e}^{-\mathrm{i}\omega t}\mathrm{d}t
+\int_{-\infty}^{0}\langle M^{x}M^{x}(t)\rangle_{\mathrm{eq}}\mathrm{e}^{-\mathrm{i}\omega t}\mathrm{d}t.
\end{eqnarray}
The second integration term can be reduced to the following form:
\begin{eqnarray}
\int_{-\infty}^{0}\langle M^{x}M^{x}(t)\rangle_{\mathrm{eq}}\mathrm{e}^{-\mathrm{i}\omega t}\mathrm{d}t
=\int_{0}^{\infty}\langle M^{x}M^{x}(-t)\rangle_{\mathrm{eq}}\mathrm{e}^{+\mathrm{i}\omega t}\mathrm{d}t
=\int_{0}^{\infty}\left(\langle M^{x}M^{x}(t)\rangle_{\mathrm{eq}}\mathrm{e}^{-\mathrm{i}\omega t}\right)^{*}\mathrm{d}t,
\end{eqnarray}
where the first equal sign follows from the change of variables $t\rightarrow -t$ and the second one the relation $\left(\langle M^{x}M^{x}(t)\rangle_{\mathrm{eq}}\right)^{*}=\langle M^{x}M^{x}(-t)\rangle_{\mathrm{eq}}$ ($*$ denote complex conjugate).
Therefore we have
\begin{eqnarray}
\int_{-\infty}^{\infty}\langle M^{x}M^{x}(t)\rangle_{\mathrm{eq}}\mathrm{e}^{-\mathrm{i}\omega t}\mathrm{d}t
&=&\int_{0}^{\infty}\left[\langle M^{x}M^{x}(t)\rangle_{\mathrm{eq}}\mathrm{e}^{-\mathrm{i}\omega t} + \left(\langle M^{x}M^{x}(t)\rangle_{\mathrm{eq}}\mathrm{e}^{-\mathrm{i}\omega t}\right)^{*}\right]\mathrm{d}t\\
&=&2\mathrm{Re}\left[\int_{0}^{\infty}\langle M^{x}M^{x}(t)\rangle_{\mathrm{eq}}\mathrm{e}^{-\mathrm{i}\omega t}\mathrm{d}t\right]
,
\end{eqnarray}
where $\mathrm{Re[\cdot]}$ denotes the real part.

The autocorrelation function $\langle M^{x}M^{x}(t)\rangle_{\mathrm{eq}}$
is given by
\beq
\langle M^{x}M^{x}(t)\rangle_{\mathrm{eq}}
={
{\rm Tr}M^{x}\mathrm{e}^{\mathrm{i}\mathcal{H}t}M^{x}
\mathrm{e}^{-\mathrm{i}\mathcal{H}t}e^{-\beta\cal H}
\over
{\rm Tr}e^{-\beta{\cal H}}
}.
\eeq
For a small system whose eigenstates can be obtained, we can calculate the trace numerically exactly. However, for larger systems we can obtain it by making use of what we call random vectors\cite{iitaka,machida,hams}. In this method we prepare the so-called Boltzmann-weighted random vectors or typical state (see below) \cite{iitaka,machida,hams,shimizu} 
$|\Phi_{\beta}\rangle$, and perform time evolution of the state by applying $e^{-i{\cal H}t}$, 
which can be done by the Chebyshev method. In this method, 
we need memory only of the order $D$ not $D^2$.
Thus we can calculate up to larger number of spins. 

Let us explain the method of the typical state briefly.
Let $\{|n\rangle\}_{n=1}^{D}$ be an arbitrary set of complete orthonormal states of Hilbert space. Using the complex-valued random variables $\{\xi_{n}\}_{n=1}^{D}$, we define the random vector as 
\begin{eqnarray}
	|v\rangle=\sum_{n=1}^{D}\xi_{n}|n\rangle.
\end{eqnarray}
The coefficients $\{\xi_{n}\}_{n}$ satisfy the following conditions:
\begin{eqnarray}
	\mathrm{E}[\xi_{n}]=0,\\
	\mathrm{E}[\xi_{m}\xi_{n}]=0,\\
	\mathrm{E}[\xi_{m}^{*}\xi_{n}]=\delta_{mn}.
\end{eqnarray}
In this paper, we draw them from the 2D dimensional spherical surface: $\sum_{n=1}^{D}|\xi_{n}|^2=D$.
This construction is independent of the choice of the set of bases $\{|n\rangle\}_{n}$.
The thermal state (or typical state) in the system Hamiltonian $\mathcal{H}$ at an inverse temperature $\beta$ is defined by
\begin{eqnarray}
	|\Phi_{\beta}\rangle=\mathrm{e}^{-\frac{1}{2}\beta\mathcal{H}}|v\rangle.
\end{eqnarray}
Using this state, we can calculate the expectation value of an observable $X$ in equilibrium by
\begin{eqnarray}            	   
       \langle X\rangle_{\mathrm{eq}}=\frac{\mathrm{Tr}[X\mathrm{e}^{-\beta\mathcal{H}}]}{\mathrm{Tr}	       
        [\mathrm{e}^{-\beta\mathcal{H}}]}=\frac{\mathrm{E}[\langle\Phi_{\beta}|X|\Phi_{\beta}\rangle]}{\mathrm{E}  
        [\langle\Phi_{\beta}|\Phi_{\beta}\rangle]}.
\end{eqnarray}
In the program, we take the average $\mathrm{E}[\cdot]$ with respect to the same samples $\{|\Phi_{\beta}\rangle\}$ in the dominator and in the numerator.
In fact, it is known that the convergence of $\mathrm{E}[\cdot]$ becomes faster with the increase of the dimension $D$ of the Hilbert space.
If we evaluate this quantitatively by using finite samples ($j=1,\cdots S$) of the states the deviation is given by\cite{hams}
\beq
{\rm P}\left(
{|{\rm Tr}[X]-{D\over S}\sum_{j=1}^S\langle \Phi_{\beta}^j|X|\Phi_{\beta}^j\rangle|^2
\over
 |{\rm Tr}[X]|^2}\ge a\right)
 \le 
 {1\over aS(D+1)}{D{\rm Tr}[X^{\dagger}X]-|{\rm Tr} [X]|^2\over |{\rm Tr} [X]|^2},\quad a>0.
 \label{Ptypical}
\eeq
So it would be enough to prepare only a few typical states as samples for a sufficiently large system.

The autocorrelation function $\langle M^{x}M^{x}(t)\rangle_{\mathrm{eq}}$ is given by
\begin{eqnarray}
	 \langle M^{x}M^{x}(t)\rangle_{\mathrm{eq}}=\frac{\mathrm{E}[\langle\Phi_{\beta}|M^{x}\mathrm{e}^{\mathrm{i}\mathcal{H}t}M^{x}\mathrm{e}^{-\mathrm{i}\mathcal{H}t}|\Phi_{\beta}\rangle]}{\mathrm{E}[\langle\Phi_{\beta}|\Phi_{\beta}\rangle]}.
\end{eqnarray}
The essence of this method is how to deal with $\mathrm{e}^{-\beta\mathcal{H}}$ and $\mathrm{e}^{-\mathrm{i}\mathcal{H}t}$. 

Here, we review this problem introducing the expansion with the Chebyshev polynomial as follows:
\begin{eqnarray}|\Phi_{\beta}\rangle&=&\mathrm{e}^{-\beta\overline{\lambda}/2}\mathrm{e}^{-\tau\mathcal{H}_{\mathrm{sc}}}|	
	v\rangle\\&=&\mathrm{e}^{-\beta\overline{\lambda}/2}\left[I_{0}(-\tau)T_{0}
	(\mathcal{H}_{\mathrm{sc}})+2\sum_{k=1}^{k_{\mathrm{max}}}I_{k}
	(-\tau)T_{k}(\mathcal{H}_{\mathrm{sc}})\right]|\Phi_{\beta}\rangle,
\end{eqnarray}
where
\begin{eqnarray}
	\mathcal{H}=\Delta\lambda\mathcal{H}_{\mathrm{sc}}+\overline{\lambda},\quad
	\Delta\lambda=\frac{E_{\mathrm{max}}-E_{\mathrm{min}}}{2},\quad	\overline{\lambda}=\frac{E_{\mathrm{max}}+
	E_{\mathrm{min}}}{2}
\end{eqnarray}
and $E_{\mathrm{max}}$ and $E_{\mathrm{min}}$ are the largest and smallest eigenvalues of $\mathcal{H}$.
The infinite sum should be truncated after $k_{\mathrm{max}}$ terms, which is chosen such that $I_{k_{\mathrm{max}}}(-\tau)$ is sufficiently small. $I_{k}(\cdot)$ denotes the modified Bessel function of order $k$ and $T_{k}(\cdot)$ the Chebyshev polynomial of order $k$. The Chebyshev polynomials satisfy the following recurrence relation:
\begin{eqnarray}
	T_{k+1}(\mathcal{H}_{\mathrm{sc}})=2\mathcal{H}_{\mathrm{sc}}T_{k}(\mathcal{H}_{\mathrm{sc}})-T_{k-1}	
	(\mathcal{H}_{\mathrm{sc}}),\quad T_{0}(\mathcal{H}_{\mathrm{sc}})=1,\quad
	T_{1}(\mathcal{H}_{\mathrm{sc}})=\mathcal{H}_{\mathrm{sc}}.
\end{eqnarray}
Using this expansion and recurrence relation, we can obtain $|\Phi_{\beta}\rangle$ only multiplying a vector by
$\mathcal{H}_{\mathrm{sc}}$ repeatedly, and summing them up, without storing any large matrices such as the Hamiltonian $\mathcal{H}$. We can also treat the time evolution operator $\mathrm{e}^{-\mathrm{i}\mathcal{H}t}$ similarly.

In this way, we obtain the time-series data of the vectors:
\beq
M^x e^{-i{\cal H}t}|\Phi_{\beta}\rangle,\quad {\rm and}\quad e^{-i{\cal H}t}M^x|\Phi_{\beta}\rangle
\eeq
and the
autocorrelation function $f(t)\equiv\langle\Phi_{\beta}|M^{x}M^{x}(t)|\Phi_{\beta}\rangle$.

\begin{figure}[H]
\vspace*{-10mm}
		{
			\includegraphics[width=100mm]{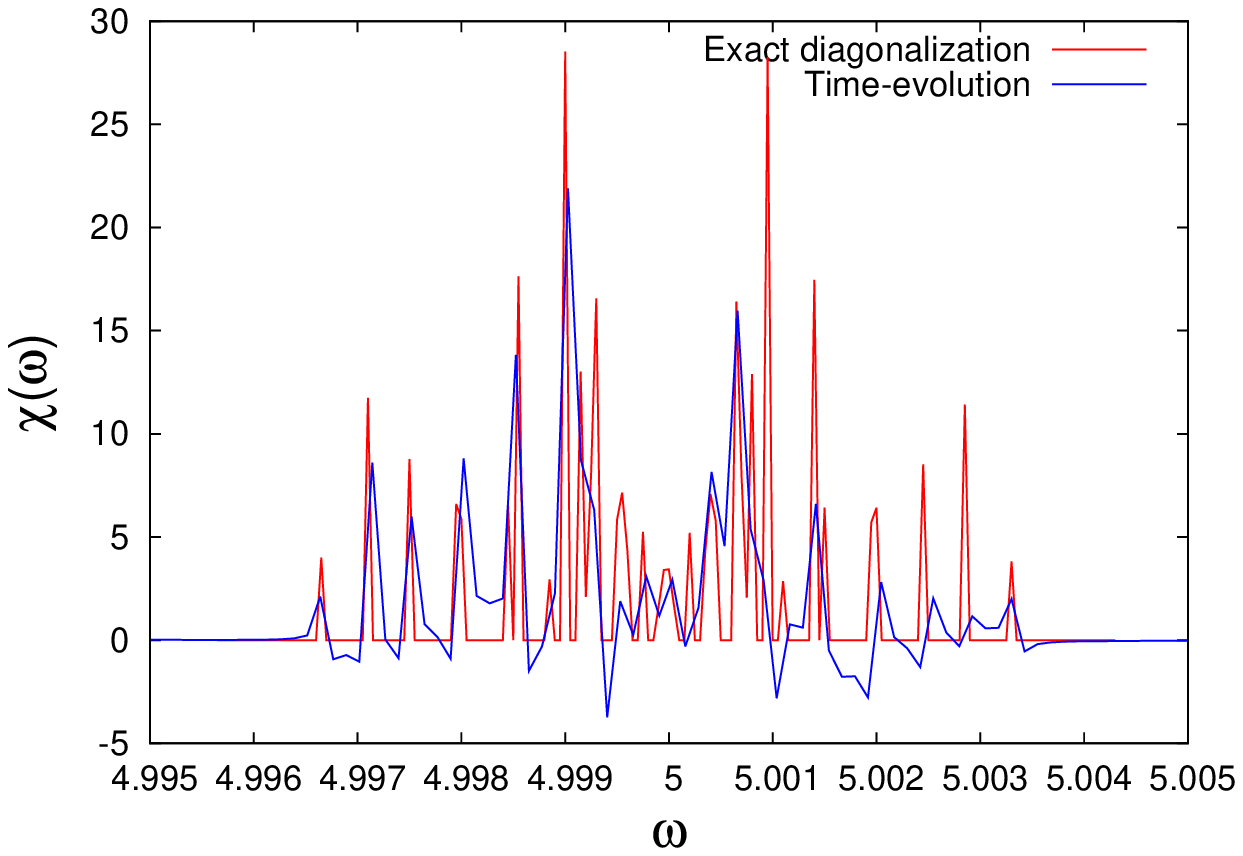}}
		{
			\includegraphics[width=100mm]{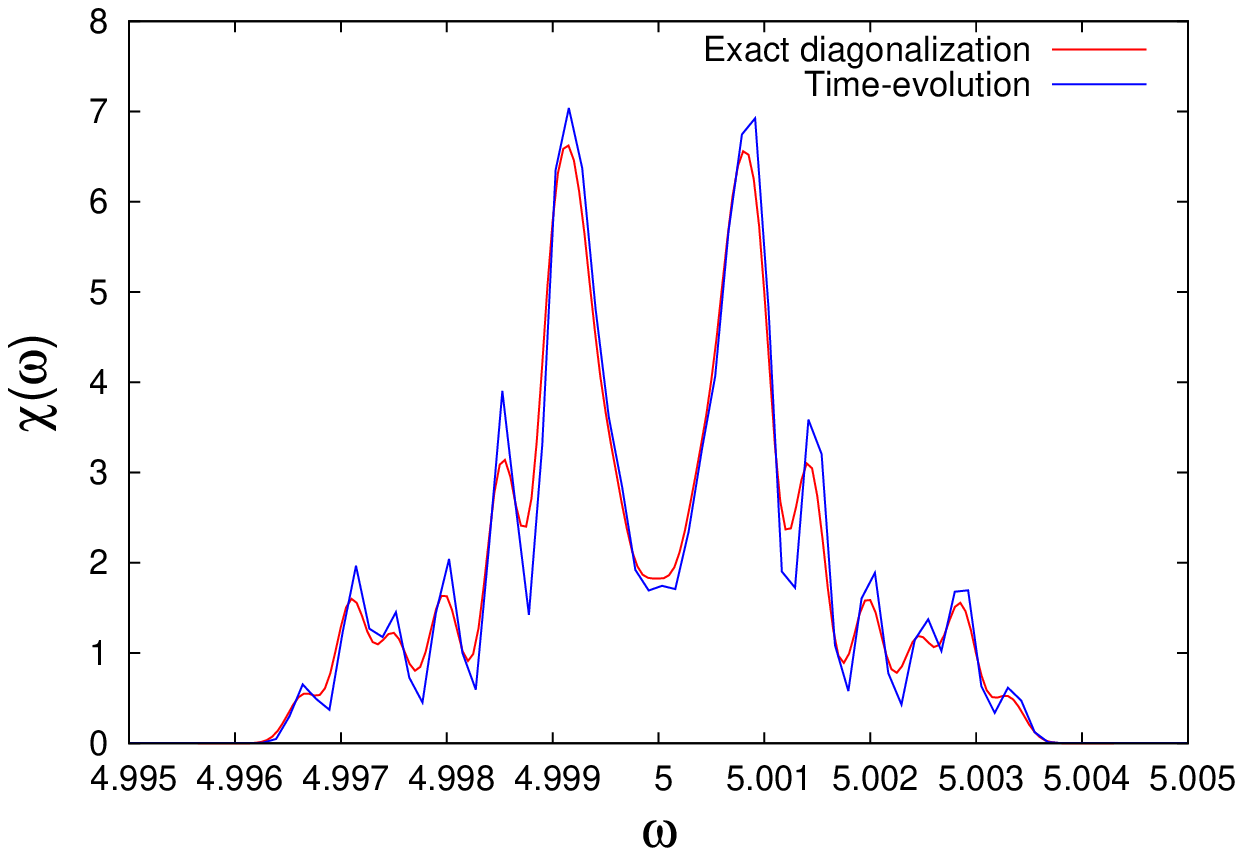}}
			
		\vspace*{20mm}
		\caption{The spectrum obtained by exact diagonalization method and by the time-evolution method : $N=6$, temperature $T=500\mathrm{K},H=\mathrm5{K}$, sampling time $\mathrm{d}t=0.5$, the number of  data $n=100000$, and the mesh of the frequency $\Delta\omega=0.00005$ (red line) and $2\pi/50000$ (blue line).
 (a) the spectrum obtained by exact diagonalization method (red line) and by time-evolution method without window functions (blue line). You can see the negativeness of the spectrum caused by the Gibbs oscillation.  (b) The smeared Gibbs oscillation owing to the Gaussian window function with $\alpha = 6$ which means that the variance in the frequency domain is $\frac{\alpha}{T}=0.00012$. Compared with the spectrum obtained by exact diagonalization with the delta peaks replaced by the Gaussians with variance = $(0.00015)^2$ (red line), the time-evolution spectrum (blue line) is consistent as expected.}
\label{EDTE}
\end{figure}

 Next, we apply discrete Fourier transform(DFT) on them. Specifically, using the set of discrete values $\{f_{k}\}_{k=0}^{n-1}\equiv\{f(\frac{kT}{n})\}_{k=0}^{n-1}$ for given $T$ and $n$, we obtain the DFT result as a set of discrete values
\begin{eqnarray}
	\int_{0}^{T}f(t)\mathrm{e}^{-\mathrm{i}\omega t}\mathrm{d}t
	\simeq\frac{T}{n}\sum_{k=0}^{n-1}f\left(\frac{kT}{n}\right)\mathrm{e}^{-\mathrm{i}\omega_{k}\frac{kT}{n}},\quad		
	\omega\simeq\omega_{k}\equiv\frac{2\pi}{T}k,\quad k=0,1,2,...,n-1,
\end{eqnarray}
where only the first half of $\{\omega_{k}\}_{k}$ are significant since the latter half correspond to the negative frequencies, and the maximum frequency is less than $\frac{\pi n}{T}$.
It should be noted that here the range of the time integral is finite but not $\infty$ as in the definition of original Fourier transform.
This leads to what is called the Gibbs oscillation problem\cite{harris} (Appendix B). 
The Gibbs oscillation with the negative values of the spectrum occurs due to the discontinuity at the ends of the interval of integration. To avoid this phenomenon, we may apply a suitable window function to DFT\cite{harris}. Here, we use a Gaussian window: 
\begin{eqnarray}
	\int_{-T}^{T}f(t)\mathrm{e}^{-\mathrm{i}\omega t}
	\mathrm{e}^{-\frac{1}{2}\left(\alpha\frac{t}{T}\right)^2}\mathrm{d}t
	=2\mathrm{Re}\left[\int_{0}^{T}f(t)\mathrm{e}^{-\mathrm{i}\omega t}
	\mathrm{e}^{-\frac{1}{2}\left(\alpha\frac{t}{T}\right)^2}\mathrm{d}t\right].
\end{eqnarray}
The Gaussian window function smears the Gibbs oscillation and gives a Gaussian-like profile with correct amplitude. As depicted in Fig.~\ref{EDTE}, the Gibbs oscillation disappears and the result is consistent with the one derived with the exact diagonalization method with the resolution of $O(\alpha)$.

\section{Wiener-Khinchin method}\label{sec_WK}

Now, we propose a new method making use of the Wiener-Khinchin theorem. 
In this method, we use the time-evolution of the state as in the previous subsection, and thus by this method we can study large systems as well.
But here, instead of the autocorrelation function, we study the dynamics of magnetization itself, i.e., $\langle M^x(t)\rangle$ from an initial state $|\Phi_{\beta}\rangle$, and use the Wiener-Khinchin relation, that is, the relation between the Fourier transform of the fluctuation in time of a quantity $X(t)$, and the spectrum of the autocorrelation function  $\langle X(0)X(t)\rangle$.

\subsection{Wiener-Khinchin theorem}
 
First, we briefly review classical Wiener-Khinchin theorem. For a time-evolution of a quantity $X(t)$,
 we define the autocorrelation function $R(t)$ and the spectral density $S(\omega)$ as
\beq
	R(t)\equiv\langle X(0)X(t)\rangle\equiv\lim_{T \to \infty}\frac{1}{T}\int_{-T/2}^{T/2}X(\tau)X(t+\tau)\mathrm{d}\tau,
	\label{Rt}
\eeq
and
\beq
	S(\omega)\equiv\lim_{T \to \infty}\frac{|X^{T}(\omega)|^2}{T},\quad X^{T}(\omega)\equiv\int_{-T/2}^{T/2}X(t)\mathrm{e}^{-\mathrm{i}\omega t}\mathrm{d}t,
	\label{Somega}
\eeq
respectively. 
The Wiener-Khinchin theorem tells us that the Fourier transform of the autocorrelation function equals to the spectral density:
\begin{eqnarray}
	G(\omega)\equiv\int_{-\infty}^{\infty}R(t)\mathrm{e}^{-\mathrm{i}\omega t}\mathrm{d}t=S(\omega).
	\label{Gomega}
\end{eqnarray}
Note that here we assumed that process $X(t)$ is stationary. 

The autocorrelation function $R(t)$ is defined as the autocorrelation function along the time-evolution. We also assume that the average is the same as that in the ensemble average in the equilibrium state if the process $X(t)$ is ergodic. 

\subsection{Dynamics of the magnetization}

Now we go back to the original model in question. We need the autocorrelation function
$R(t)=\langle M^{x}(0)M^{x}(t) \rangle_{\mathrm{eq}}$ to obtain the spectral density $S(\omega)$ which is directly related to the ESR spectrum. 
Here it should be noted that $M^{x}(t)$ is an operator and the definition of $\langle M^{x}(t)\rangle$ is tricky. Definitely, $\langle M^{x}(t)\rangle_{\mathrm{eq}}$ is time-independent. In the following, we propose a numerical method motivated by the Wiener-Khinchin theorem, but it should be understood as a numerical algorithm to obtain $S(\omega)$.\cite{discuss} 

First we prepare a typical initial state for the canonical ensemble,
$|\Phi_{\beta}\rangle=e^{-\beta{\cal H}/2}|v\rangle$. The expectation value with the state gives the equilibrium state at the given temperature. But for an sample of the typical state,
 $\langle M^{x}(t)\rangle$ is not necessarily zero.
Then we calculate the time evolution of the selected state and obtain the value of 
$\langle M^{x}(t)\rangle$ at each time. From this time evolution of $\langle M^{x}(t)\rangle$, we obtain the spectral density Eq.(\ref{Somega}). 

At last, by averaging the spectral densities with respect to several initial states,
we obtain the spectral density for this process.
Here we again note that the expectation value $\mathrm{E}[\langle M^{x}(t)\rangle]$ in the equilibrium is zero.  Thus, the expectation value of the spectral density of its Fourier transform
\beq
M_{\beta}^{T}(\omega)\equiv \int_{T/2}^{T/2}\langle M^{x}(t)\rangle\mathrm{e}^{-\mathrm{i}\omega t}\mathrm{d}t
\eeq
is also zero.

However, the expectation value of the spectral density 
$\mathrm{E}[|M_{\beta}^{T}(\omega)|^2]$ is not zero, and this provides information for 
the Fourier transform of the autocorrelation function $G(\omega)$ for the ESR spectrum of the system.
We will show below the explicit relation of this spectral density and $G(\omega)$.


\subsection{Relation between the spectrum density and $G(\omega)$}
The time evolution of the system from a given initial state $|\Phi_{\beta}\rangle$ provided with the random coefficients $\{\xi_n\}$ is assumed to be given by the quantum mechanical evolution 
$\mathrm{e}^{-\mathrm{i}\mathcal{H}t}$.

Thus, we introduce a quantity ${\hat M}_{\beta}^{x}(t)$ by
\begin{eqnarray}
	{\hat M}_{\beta}^{x}(t)\equiv\langle\Phi_{\beta}|M^{x}(t)|\Phi_{\beta}\rangle
	=\sum_{m,n}\xi_{m}^{*}\xi_{n}\mathrm{e}^{-\beta(E_{m}+E_{n})/2}\mathrm{e}^{\mathrm{i}(E_{m}-E_{n})t}
	\langle m|M^{x}|n\rangle,
	\label{Mbetat}
\end{eqnarray}
where 
$|n\rangle$ and $E_n$ are the eigenvector and its eigenenergy of the system Hamiltonian ${\cal H}$.

The Fourier transform ${\hat M}_{\beta}^{T}(\omega)$ is expressed as
\begin{eqnarray}
	{\hat M}_{\beta}^{T}(\omega)	=\sum_{m,n}\xi_{m}^{*}\xi_{n}\mathrm{e}^{-\beta(E_{m}+E_{n})/2}2\pi\delta^{T}\left(\omega-(E_{m}-E_{n})\right)
	\langle m|M^{x}|n\rangle,
	\label{Mbetaomega}
\end{eqnarray}
where
\begin{eqnarray}
\delta^{T}(\omega)\equiv\frac{1}{2\pi}\int_{-T/2}^{T/2}\mathrm{e}^{-\mathrm{i}\omega t}\mathrm{d}t=\frac{\mathrm{sin\frac{\omega T}{2}}}{\pi\omega}\xrightarrow{T \to \infty}\delta(\omega).
\end{eqnarray}
Therefore
\begin{multline}
	|{\hat M}_{\beta}^{T}(\omega)|^2=\sum_{m,n}\sum_{m',n'}\xi_{m}^{*}\xi_{n}\xi_{m'}\xi_{n'}^{*}
	\mathrm{e}^{-\beta(E_{m}+E_{n})/2}\mathrm{e}^{-\beta(E_{m'}+E_{n'})/2}\\
	\times4\pi^{2}\delta^{T}\left(\omega-(E_{m}-E_{n})\right)\delta^{T}\left(\omega-(E_{m'}-E_{n'})\right)
	\langle m|M^{x}|n\rangle\langle n'|M^{x}|m'\rangle.
	\label{Zsquare}
\end{multline}
Here we take the average over the random variables $\{\xi_{n}\}_{n}$. Using the following formula\cite{hams}:
\begin{eqnarray}
	\mathrm{E}[\xi_{m}^{*}\xi_{n}\xi_{m'}\xi_{n'}^{*}]=\frac{D}{D+1}\{\delta_{m,n}\delta_{m',n'}	 
       (1-\delta_{m,m'})+\delta_{m,m'}\delta_{n,n'}(1-\delta_{m,n})\}
	+\frac{2D}{D+1}\delta_{m,n'}\delta_{n,m'}\delta_{m,n},
\end{eqnarray}
and $\langle n|M^{x}|n\rangle$=0, we have
$$
	\mathrm{E}[|{\hat M}_{\beta}^{T}(\omega)|^2]=\sum_{m,n}\mathrm{E}[|\xi_{n}|	     
       ^2|\xi_{m}|^2]\mathrm{e}^{-\beta(E_{m}+E_{n})}4\pi^{2}
	\left(\delta^{T}(\omega-(E_{m}-E_{n}))\right)^2|\langle m|M^{x}|n\rangle|^2
	$$
\begin{eqnarray}	
	=\frac{D}{D+1}T\mathrm{e}^{-\beta\omega}\sum_{m,n}\mathrm{e}^{-2\beta E_{n}}2\pi
	\delta^{T}\left(\omega-(E_{m}-E_{n})\right)|\langle m|M^{x}|n\rangle|^2,
	\label{EMbetaomega}
\end{eqnarray}
where we used the relation
\begin{eqnarray}
	\left(\delta^{T}(\omega-(E_{m}-E_{n}))\right)^2\approx\frac{T}{2\pi}\delta^{T}
	\left(\omega-(E_{m}-E_{n})\right),
\end{eqnarray}
and the final result for the spectral density reads
\begin{eqnarray}
	\Sigma_{\beta}(\omega)&=&\lim_{T \to \infty}\frac{\mathrm{E}[|{\hat M}_{\beta}^{T}(\omega)|^2]/Z_{\beta}^2}{T}\\
	&=&\frac{D}{D+1}\mathrm{e}^{-\beta\omega}\sum_{m,n}\mathrm{e}^{-2\beta E_{n}}2\pi
	\delta\left(\omega-(E_{m}-E_{n})\right)|\langle m|M^{x}|n\rangle|^2/Z_{\beta}^2,	
\end{eqnarray}
where
\beq
Z_{\beta}=\mathrm{E}[\langle \Phi{_\beta}| \Phi{_\beta}\rangle].
\eeq

Finally by comparing this result with the Fourier transform of the autocorrelation function
\begin{eqnarray}
	G_{\beta}(\omega)=\sum_{m,n}\mathrm{e}^{-\beta E_{n}}2\pi
	\delta\left(\omega-(E_{m}-E_{n})\right)|\langle m|M^{x}|n\rangle|^2/Z_{\beta},
\end{eqnarray}
we obtain a Wiener-Khinchin-like relation
\begin{eqnarray}
	G_{\beta}(\omega)=\frac{D+1}{D}\frac{Z_{\beta/2}^2}{Z_{\beta}}
	\mathrm{e}^{\frac{\beta\omega}{2}}\Sigma_{\beta/2}(\omega).
\end{eqnarray}
Thus, the imaginary part of the dynamical susceptibility of our interest is given by
\begin{eqnarray}
	\chi''(\omega)=\frac{D+1}{D}\frac{Z_{\beta/2}^2}{Z_{\beta}}
	\mathrm{sinh}\left(\frac{\beta\omega}{2}\right)\Sigma_{\beta/2}(\omega).
\end{eqnarray}
Here it should be noted that we need to calculate the quantities of $\beta/2$ (not $\beta)$ to obtain the ESR spectrum of $\beta$. 

In the present method, we obtained the approximate spectrum from a finite time domain. The finiteness of the domain causes the Gibbs oscillation in this case as well.
However, because we take the square of $\hat{M}_{\beta}^{T}(\omega)$ (Eq.(\ref{Zsquare})), and thus the negative value would not appear. 

Here we note that the total amplitude of the spectrum, i.e., the integration over the
spectrum does not change because of the relation:
\beq
{1\over T}\int_{-\infty}^{\infty}\left( {\sin\omega T/2\over \pi\omega}\right)^2\mathrm{d}\omega=\frac{1}{2\pi},
\label{H2}
\eeq
and we can obtain the correct spectrum in the limit $T\rightarrow\infty$ in the 
present method.

As to the effect of the Gibbs oscillation,
in the present method it gives the width of the spectrum due to the finiteness of the observation. 
In this sense we regard that the width comes from the way of observation and is natural, in contrast to the case of the method presented in the section \ref{sec:TDACFN} where we introduced a gaussian window with an artificial width to smear the Gibbs oscillation which is the same order of the natural width. The width of the gaussian window is taken to be in the same order of the width of the Gibbs oscillation.

We show a comparison of the spectrum obtained by the exact diagonalization method and by the Wiener-Khinchin method. 
\begin{figure}[H]
	\begin{center}
	\includegraphics[width=100mm]{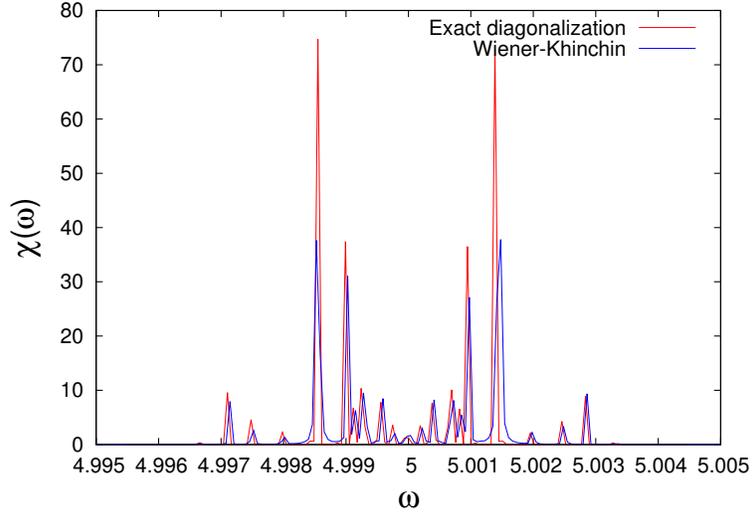}
	\end{center}
	\vspace{10mm}
	\caption{Absorption spectrum obtained by the exact diagonalization method and by the Wiener-Khinchin method: $N=6$ (100 samples), $T=200\mathrm{K}$, $H=5\mathrm{K}$, and the mesh of the frequency $\Delta\omega=0.000063$ (red line) and $2\pi/100000$ (blue line).}
	\label{wiener-khinchin}
\end{figure}

We also show the intensity 
\beq
I^x=\int_0^{\infty} I^x(\omega)\mathrm{d}\omega
\eeq
 which is the integral over the spectrum $I^x(\omega)$ obtained by the exact diagonalization method  and  by the Wiener-Khinchin method as a function of temperature. The data of WK method are obtained by averaging 100 samples. Here we find almost perfect agreement.
\begin{figure}[H]
	\begin{center}
	\includegraphics[width=100mm]{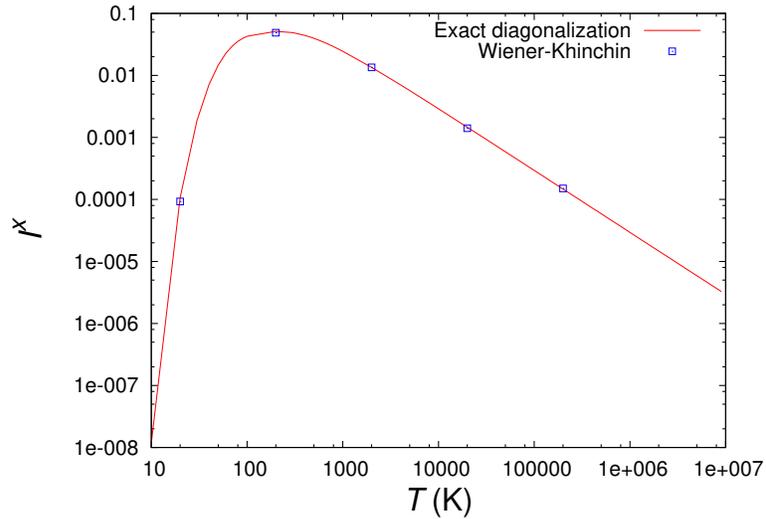}
	\end{center}
	\vspace{10mm}
	\caption{Comparison of intensities of the absorption as a function of the temperature: $N=6$,$H=5\mathrm{K}$, the number of samples $S=100$.}
	\label{intensity}
\end{figure}

\subsection{Typicality of the fluctuation}

If ${\rm Tr}[X]\neq 0$, the estimation Eq.(\ref{Ptypical}) works. 
But, in WK method ${\rm Tr} [Xe^{-\beta{\cal H}}]= 0$, and thus we cannot use this bound for the estimation of the error
for
the quantity ${\hat M}_{\beta}^{x}(t)$ (Eq.~(\ref{Mbetat})) or ${\hat M}_{\beta}^{x}(\omega)$  (Eq.~(\ref{Mbetaomega})). Thus to obtain
$\mathrm{E}[|{\hat M}_{\beta}^{T}(\omega)|^2]$  (Eq.~(\ref{EMbetaomega})) from numerical data,
we need to estimate the variance of the quantity:
\beq
\mathrm{E}\left[\left( |{\hat M}_{\beta}^{T}(\omega)|^2-\mathrm{E}[|{\hat M}_{\beta}^{T}(\omega)|^2]\right)^2\right]
=\mathrm{E}[
|{\hat M}_{\beta}^{T}(\omega)|^4] - \mathrm{E}[|{\hat M}_{\beta}^{T}(\omega)|^2]^2
\eeq
If we assume the non-degeneracy of energy gaps that if $E_{m}-E_{n}=E_{m'}-E_{n'}$ then ($m=m'$ and $n=n'$) or ($m=n$ and $m'=n'$),
and approximate $\delta^T(x)=\delta(x)$, then
we have
\beq
\mathrm{E}\left[\xi_m^*\xi_n\xi_{m'}\xi_{n'}^*\xi_k^*\xi_l\xi_{k'}\xi_{l'}^*\right]
\rightarrow\mathrm{E}\left[(\xi_m^*\xi_m)^2(\xi_{n}^*\xi_{n})^2\right]
\simeq 4,
\eeq
and 
\beq 
\mathrm{E}[
|{\hat M}_{\beta}^{T}(\omega)|^4] - \mathrm{E}[|{\hat M}_{\beta}^{T}(\omega)|^2]^2\simeq 3\mathrm{E}[|{\hat M}_{\beta}^{T}(\omega)|^2]^2.
\eeq
Thus, the distribution of the quantity $|{\hat M}_{\beta}^{T}(\omega)|^2$ has a standard deviation
\beq
\sqrt{3}\mathrm{E}[|{\hat M}_{\beta}^{T}(\omega)|^2],
\eeq
which does not depend on $D$ and of the order of the value of average.
This means that the distribution function converges to a fixed form. 
Thus, we can estimate $\mathrm{E} [|{\hat M}_{\beta}^{T}(\omega)|^2]$ by  some ensemble average over the sample independently of $D$.

In the case of the autocorrelation function ${\rm Tr}[Xe^{-\beta{\cal H}}]\ne 0$, and we expect that the distribution of the obtained values converges to the expectation value with the variance of the order $1/D$.
In this sense, the method of the autocorrelation has an advantage. But the WK method can also give a value with finite sampling because the variance of which does not depend on $D$, and the distribution  shows a kind of typicality\cite{paper}. We may call it the typicality of the distribution for zero-mean quantity.

\section{Summary and Discussion}\label{sec_summary}

We have proposed a time-domain method to obtain ESR spectrum by making use of Wiener-Khinchin relation, in which effects of the Gibbs oscillation due to finite observation period is practically suppressed. This method is used the thermal typical state for the variable whose expectation value is zero in the thermal average, and thus the temperature of the thermal state and that of the spectrum differ by the factor 2. 

Here we presented one of the way for time-domain method motivated by the Wiener-Khinchin theorem, but there are many other ways to obtain the spectrum density by making use of relation related to Wiener-Khinchin theorem.
For example, if we calculate ${\rm E}[\langle\Phi_{\beta}|M^x|\Phi_{\beta}\rangle\langle\Phi_{\beta}|M^x(t)|\Phi_{\beta}\rangle]$, we can obtain a similar expression. The relation among them will be interesting problem.

It should be noted that the present method does not directly relate to the experimental situation. 
In this paper, we proposed the present method by making use of mathematical relations discussed in Sec \ref{sec_WK} to obtain the spectrum. But, dynamics from the thermal typical state is an interesting problem to study, which will be studied elsewhere.



\begin{acknowledgements}
The authors thank Professor Hans De Raedt for his very valuable and stimulating comments. 
The present work was supported by 
Grants-in-Aid for Scientific Research C (25400391) from MEXT of Japan, and the Elements 
Strategy Initiative Center for Magnetic Materials under the 
outsourcing project of MEXT. The numerical calculations were supported by the supercomputer center of ISSP of Tokyo University.
We also acknowledge the JSPS Core-to-Core Program: Non-equilibrium dynamics of soft matter 
and information.
\end{acknowledgements}

\appendix 

\section{Model for Application}
We demonstrate the methods applying to the one-dimensional spin-1/2 XXZ model with the dipole-dipole interaction under a static magnetic field $H$ and an oscillating field $\lambda(t)$. We obtain the response to the AC field $\lambda(t)$ along the $x$ axis. The hamiltonian of the system is given by
\begin{eqnarray}
	\mathcal{H_{\mathrm{tot}}}&=&\mathcal{H} + \lambda(t)\\
	&=&\mathcal{H}_{0} + \mathcal{H'} + \mathcal{H}_{\mathrm{D}} + \mathcal{H}_{\mathrm{Z}} +\lambda(t), 
		\label{eq:XXZ}	
\end{eqnarray}
where
\begin{eqnarray}
	\mathcal{H}_0&=&J\sum_{i=1}^{N-1}\mbox{\boldmath $S$}_{i}\cdot\mbox{\boldmath $S$}_{i+1},\\
	\mathcal{H'}&=&J\sum_{i=1}^{N-1}\Delta S_{i}^{z}S_{i+1}^{z},\\
	\mathcal{H}_{\mathrm{D}}&=&D_{0}\sum_{\langle i,j\rangle}\left(\frac{\mbox{\boldmath $S$}_{i}\cdot\mbox{\boldmath $S$}_{j}}{r_{ij}^{3}}-\frac{3(\mbox{\boldmath $S$}_{i}\cdot\mbox{\boldmath $r$}_{ij})(\mbox{\boldmath $S$}_{j}\cdot\mbox{\boldmath $r$}_{ij})}{r_{ij}^{5}}\right),\\
		\label{eq:Hz}
	\mathcal{H}_{\mathrm{Z}}&=&-g\mu_{B}H\sum_{i=1}^{N}S_{i}^{z},\\
	\lambda(t)&=&\lambda_{0}\mathrm{cos}\omega t\sum_{i=1}^{N}S_{i}^{x}.
\end{eqnarray}
Hereafter we put  $g\mu_{B}=1$. We adopt Kelvin as the unit of enegy, and $\Delta=0.00001$.
For the demonstration we put $D_{0}=0$.

\section{Gibbs oscillation and window function}
\label{sec:appendixA}
Here we discuss the properties of Gibbs oscillation which inevitably comes from the finite observation time.  This is a general problem encountered in performing the Fourier transform of time-series data in finite time domain\cite{harris}.
In the method of autocorrelation function this effect causes undesirable negative values of the spectrum, which has been managed to be smeared by the use of a window function.

\begin{figure}[H]
	\begin{center}
	\includegraphics[width=100mm]{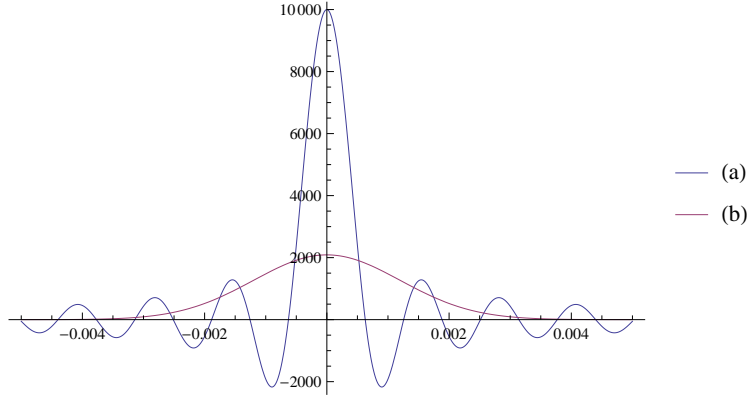}
	\end{center}
	\caption{Square root of power spectrum of window function : $n=10000$, $\Delta=0.5$, and $\alpha=6$.  (a) rectangle window, (b) gaussain window.}
	\label{window}
\end{figure}

First let us have a brief review on the Gibbs oscillation.
Let $\{f_{k}\}_{k=-\infty}^{\infty}$ be the time sequence of data such as the autocorrelation function. The spectrum of process is given by
\begin{eqnarray}
	F(\omega)&=&\Delta\sum_{k=-\infty}^{\infty}f_{k}\mathrm{e}^{-\mathrm{i}\omega k 
	\Delta}\left(\simeq\int_{-\infty}^{\infty}f(t)\mathrm{e}^{-\mathrm{i}\omega t}\mathrm{d}t\right),\\
	f_{k}&=&\frac{1}{2\pi}\int_{-\pi/\Delta}^{\pi/\Delta}F(\omega)\mathrm{e}^{\mathrm{i}\omega k 	
	\Delta}\mathrm{d}\omega,
\end{eqnarray}
where $\Delta$ is the interval of sampling. This transformation is often called the discrete time Fourier transform (DTFT). In reality, however, since we cannot prepare an infinite number of data, we have to terminate the sum at a finite number. Thus the transformation is modified as
\beq
	\tilde{F}(\omega)\equiv\Delta\sum_{k=-n}^{n-1}f_{k}\mathrm{e}^{-\mathrm{i}\omega 
	k\Delta}=\Delta\sum_{k=-\infty}^{\infty}f_{k}h_{k}\mathrm{e}^{-\mathrm{i}\omega k\Delta}
\eeq
where we used a symmetry of $f_k$
\begin{eqnarray}
	h_{k}=\left\{ \begin{array}{ll}
	1 & \hspace{1cm}k=0,\pm1,\pm2...,\pm(n-1),-n\\
	0 & \hspace{1cm}\mathrm{otherwise.}\\
	 \end{array} \right.
\end{eqnarray}
This formula is expressed in the form of convolution: 
\beq
	\tilde{F}(\omega)=\frac{1}{2\pi}\int_{-\pi/\Delta}^{\pi/\Delta}F(\omega')H(\omega-\omega')
	=\frac{1}{2\pi}\int_{-\pi/\Delta}^{\pi/\Delta}F(\omega-\omega')H(\omega'),
\eeq
where $H(\omega)$ is the DFTT of $\{h_{k}\}_{k}$
\begin{eqnarray}
	{H}(\omega)=\Delta\sum_{k=-n}^{n-1}\mathrm{e}^{-\mathrm{i}\omega k\Delta}
	=\Delta \mathrm{e}^{\mathrm{i}\frac{\omega\Delta}{2}}\frac{\mathrm{sin}(\omega n\Delta)}
	{\mathrm{sin}(\frac{\omega\Delta}{2})}.
\end{eqnarray}
Above formula means that the spectrum we need is deformed by the DTFT of the rectangular window $H(\omega)$.
In Fig.~\ref{window}(a),  $\hat{H}(\omega)\equiv\Delta\frac{\mathrm{sin}(\omega n\Delta)}{\mathrm{sin}(\omega\Delta/2)}$ is depicted. This oscillation gives the negative peak of the spectrum $\tilde{F}(\omega)$. (Note that the frequency $\omega$ is also discretized in the finite domain.)

In order to avoid this apparent negative peaks, a window function method unusually with a Gaussian window has been introduced. 
\begin{eqnarray}
	g_{k}=h_{k}\mathrm{e}^{-\frac{1}{2}(\alpha\frac{k}{n})^2},\quad k=0,\pm1,\pm2,...
\end{eqnarray}
In Fig.~\ref{window}(b), the spectrum of $|G(\omega)|$ is depicted. 

The parameter $\alpha$ determines the artificial resolution of the spectrum. Within this resolution, the Gibbs oscillation is smeared out and, so we can reproduce the spectrum as discussed in section \ref{sec:TDACFN}.

Here, we take the value of $\Delta$ which is the interval of observation time arbitrary. 
But, in the case the eigenenergy is confined in a finite range $E_{\rm min}<E_i<E_{\rm max}$, that is the situation so-called Bond-limited function, the Nyquist-Shannon  theorem  tells us that $\Delta$ should be smaller than $\pi/E_{\rm max}$.
Thus, it is most efficient to take $\Delta$ in the Chebyshev procedure to be this value.

The width of the Gauss window $\alpha$ is known to be taken as 
\beq
e^{-{1\over2}\left(\alpha{n\over n}\right)^2}= \varepsilon \rightarrow \alpha^2=-2\ln \varepsilon, 
\eeq
where $\varepsilon$ is a number of the order of the smallest number of the computer resolution,
say  $\varepsilon=10^{-12}$.
This is consistent with the above choice of $\alpha=6$.
The period of the Gibbs oscillation is of the same order of the width of the gaussian window.

\begin{figure}[H]
	\begin{center}
	\includegraphics[width=100mm]{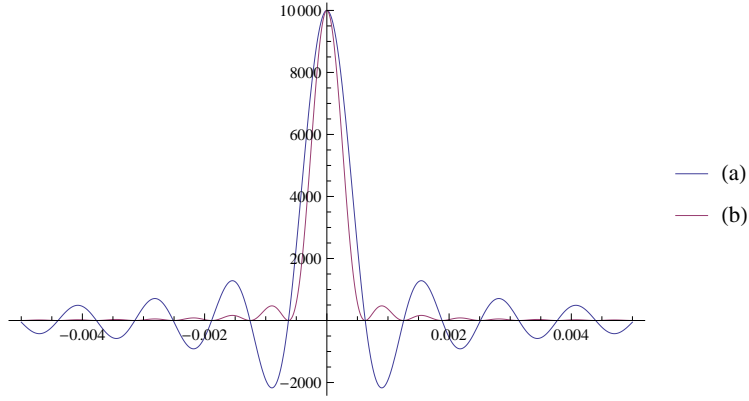}
	\end{center}
	\caption{Comparison of windows : $n=10000$, $\Delta=0.5$ : (a) $\hat{H}(\omega)$, (b)$\frac{1}{2T}\hat{H}(\omega)^2$.}
	\label{hikaku_window}
\end{figure}

In contrast to the method using the autocorrelation function, in the WK method the spectrum is squared, and the spectrum is given by $|\tilde{F}(\omega)|^2$ and thus the spectrum is positive. Although the effect of the Gibbs oscillation exists, 
it gives natural width due to the finite time window and do not harms the spectrum much.  
In Fig.~\ref{hikaku_window}, we show comparison of $\hat{H}(\omega)$ and $|\hat{H}(\omega)|^2$


\begin{thebibliography}{99}
\bibitem{R1}	M. Nishino, H. Onishi, K. Yamaguchi, and S. Miyashita, Phys. Rev. B  \textbf{62}, 9463 (2000).
\bibitem{R2}		M. Nishino, H. Onishi, P. Roos, K. Yamaguchi, and S. Miyashita, Phys. Rev. B  \textbf{61}, 4033 (2000).
\bibitem{R3}	J. Sirker, N. Laflorencie, S. Fujimoto, S. Eggert, and I. Affleck, Phys. Rev. Lett.  \textbf{98}, 137205 (2007).
\bibitem{R4}		K. Kojima, J. Yamanobe, H. Eisaki, S. Uchida, Y. Fudamoto, I. Gat, M. Larkin, A. Savici, Y. Uemura, P. Kyriakou, M. Rovers, and G. Luke, Phys. Rev. B  \textbf{70}, 094402 (2004).
\bibitem{R5}		A. I. Smirnov, V. N. Glazkov, L. I. Leonyuk, A. G. Vetkin, and R. M. Eremina, J. Exp. Theor. Phys.  \textbf{87}, 1019 (1998).
\bibitem{R6}		S. Bertaina, C.-E. Dutoit, J. Van Tol, M. Dressel, B. Barbara, and A. Stepanov, Phys. Rev. B  \textbf{90}, 060404 (2014).
\bibitem{R7}		G. De Chiara, S. Montangero, P. Calabrese, and R. Fazio, 1 (n.d.).
\bibitem{R8}		L. Campos Venuti, C. Degli Esposti Boschi, and M. Roncaglia, Phys. Rev. Lett.  \textbf{96}, 247206 (2006).
\bibitem{R9}	L. Campos Venuti, C. Degli Esposti Boschi, and M. Roncaglia, Phys. Rev. Lett.  \textbf{99}, 060401 (2007).
\bibitem{R10}		S. Bose, Phys. Rev. Lett.  \textbf{91}, 207901 (2003). 
\bibitem{miyashita}
S. Miyashita, T. Yoshino and A. Ogasahara, J. Phys. Soc. Jpn. \textbf{68}, 655-661 (1999).
\bibitem{cepas}
S. El Shawish,O. Cepas and S.Miyashita,
Phys. Rev. B  \textbf{81}, 224421 (2010).
\bibitem{iitaka}
T. Iitaka and T. Ebisuzaki, Phys. Rev. Lett. \textbf{90}, 047203 (2003).
\bibitem{machida}
M. Machida, T. Iitaka, and S. Miyashita, Phys. Rev. B \textbf{86}, 224412 (2012) 
\bibitem{hams}
A. Hams and H. De Raedt, Phys. Rev. E \textbf{62}, 4365 (2000)
\bibitem{shimizu} S. Sugiura and A. Shimizu, Phys. Rev. Lett. \textbf{108} 240401 (2012).
\bibitem{kubo_tomita}
R. Kubo and K. Tomita, J. Phys. Soc. Jpn. \textbf{9} 888 (1954).
\bibitem{kubo}
R. Kubo, J. Phys. Soc. Jpn. \textbf{12} 570 (1957).
\bibitem{harris}
F. J. Harris, in Proc, IEEE, \textbf{66}, 51 (1978).
\bibitem{discuss} The relation to the real measurement is an very interesting problem but it would be discussed separately.
\bibitem{paper} Detailed comparison in large systems will be reported:
H. Ikeuchi, S. Bertaina, H. De Raedt, and S. Miyashita: in preparation.
\end{thebibliography}
\end{document}